\documentstyle[preprint,eqsecnum,aps,english,epsfig]{revtex}

\begin{document}
\draft
\tighten		
\preprint{GSI-Preprint-98-06}		

\title{Temperatures of Exploding Nuclei}

\author
{
V.~Serfling,$^{(1)}$
C.~Schwarz,$^{(1)}$
R.~Bassini,$^{(2)}$
M.~Begemann-Blaich,$^{(1)}$
S.~Fritz,$^{(1)}$
S.J.~Gaff,$^{(3)}$
C.~Gro\ss,$^{(1)}$
G.~Imm\'{e},$^{(4)}$
I.~Iori,$^{(2)}$
U.~Kleinevo\ss,$^{(1)}$
G.J.~Kunde,$^{(3)}$\cite{AAA}   
W.D.~Kunze,$^{(1)}$
U.~Lynen,$^{(1)}$
V.~Maddalena,$^{(4)}$\cite{BBB}		
M.~Mahi,$^{(1)}$
T.~M\"ohlenkamp,$^{(5)}$\cite{BBB}
A.~Moroni,$^{(2)}$
W.F.J.~M\"uller,$^{(1)}$
C.~Nociforo,$^{(4)}$       		
B.~Ocker,$^{(6)}$
T.~Odeh,$^{(1)}$
F.~Petruzzelli,$^{(2)}$
J.~Pochodzalla,$^{(7)}$
G.~Raciti,$^{(4)}$
G.~Riccobene,$^{(4)}$			
F.P.~Romano,$^{(4)}$    		
A.~Saija,$^{(4)}$     			
M.~Schnittker,$^{(1)}$
A.~Sch\"uttauf,$^{(6)}$
W.~Seidel,$^{(5)}$
C.~Sfienti,$^{(4)}$			
W.~Trautmann,$^{(1)}$
A.~Trzcinski,$^{(8)}$
G.~Verde,$^{(4)}$
A.~W\"orner,$^{(1)}$
Hongfei~Xi,$^{(1)}$\cite{BBB}
and B.~Zwieglinski$^{(8)}$
}

\address
{
$^{(1)}$Gesellschaft  f\"ur  Schwerionenforschung, D-64291 Darmstadt, 
Germany\\
$^{(2)}$Istituto di Scienze Fisiche, Universit\`{a} degli Studi
di Milano and I.N.F.N., I-20133 Milano, Italy\\
$^{(3)}$Department of Physics and
Astronomy and National Superconducting Cyclotron Laboratory,
Michigan State University, East Lansing, MI 48824, USA\\
$^{(4)}$Dipartimento di Fisica dell' Universit\`{a}
and I.N.F.N.,
I-95129 Catania, Italy\\
$^{(5)}$Forschungszentrum Rossendorf, D-01314 Dresden, Germany\\
$^{(6)}$Institut f\"ur Kernphysik,
Universit\"at Frankfurt, D-60486 Frankfurt, Germany\\
$^{(7)}$Max-Planck-Institut f\"ur Kernphysik,
D-69117 Heidelberg, Germany\\
$^{(8)}$Soltan Institute for Nuclear Studies,
00-681 Warsaw, Hoza 69, Poland
}
\date{\today}
\maketitle

\begin{abstract}
Breakup temperatures in central collisions of
$^{197}$Au + $^{197}$Au at bombarding energies $E/A$ = 50 to 200
MeV were determined with two methods.
Isotope temperatures, deduced from double ratios of hydrogen, helium, 
and lithium isotopic yields, increase monotonically with bombarding
energy from 5 MeV to 12 MeV, in qualitative agreement with
a scenario of chemical freeze-out after adiabatic expansion.
Excited-state temperatures, derived from
yield ratios of states in $^{4}$He, $^{5,6}$Li,
and $^{8}$Be, are about 5 MeV, 
independent of the projectile energy, and
seem to reflect the internal temperature of 
fragments at their final separation from the system.
\end{abstract}


\narrowtext

Recently, a caloric curve of nuclei has been obtained by correlating
the values of temperature and excitation energy measured for spectator
fragmentation in reactions of $^{197}$Au + $^{197}$Au at 600 MeV per
nucleon \cite{poch95}. The temperatures were derived from double 
ratios of helium and lithium isotopic yields while the excitation energies
were obtained by adding up the kinetic energies of the product nuclei
and the mass excess of the observed partition with respect to the ground 
state of the reconstructed spectator nucleus. 
The double-bended shape of the caloric curve and its similarity
to predictions of microscopic statistical models 
\cite{gross90,bond95,hongfei}, has stimulated
considerable experimental and theoretical activities.
In particular, the second rise of the temperature to values exceeding
10 MeV at high excitation energies 
has initiated the discussion of whether nuclear temperatures
of this magnitude can be measured reliably 
(see Refs. \cite{hongfei,xi96,gulm97} 
and references given in these recent papers)
and whether this observation may indeed be linked to 
a transition towards the vapor phase 
\cite{natowitz,more96}. Obviously, 
a well-founded understanding of the significance of the employed
temperature observables \cite{mori94}
is indispensable when searching for signals of
the predicted liquid-gas phase transition in nuclear matter.

Here, we present the results of temperature measurements for 
central collisions of $^{197}$Au + $^{197}$Au at incident energies 
$E/A$ = 50 MeV to 200 MeV. 
These collisions are characterized by a collective radial
flow of light particles and fragments 
which, over the covered energy range, evolves as a dynamical phenomenon
closely connected to the initial stages of the reaction
\cite{reis97}.
Global equilibrium is clearly not achieved.
If local equilibrium is reached, the associated temperatures should
reflect the adiabatic cooling of the rapidly expanding system.

Two temperature observables were used simultaneously.
Isotope temperatures were deduced from double ratios of isotopic
yields \cite{albergo} and excited-state temperatures 
were derived from the correlated yields of light-particle coincidences
\cite{mori94,poch87,kunde91}.
It will become evident from the diverging results that this represents 
more than a methodical test and that the two types of thermometers
are sensitive to different stages of the fragment formation and emission.

Beams of $^{197}$Au with $E/A$ = 50, 100, 150, and 200 MeV, provided by 
the heavy-ion synchrotron SIS, were directed onto targets of 75-mg/cm$^2$ 
areal density.
Two multi-detector hodoscopes, consisting of 96 and of 64 
Si-CsI(Tl) telescopes in closely-packed geometries, were placed
on opposite sides with respect to the beam axis. 
Four high-resolution telescopes \cite{hongfei}
were used to measure the isotopically resolved yields of 
light charged particles and fragments.
The choice of angles, 
between $\theta_{lab}$ = 24$^{\circ}$ and 58$^{\circ}$ for the hodoscopes
and $\theta_{lab} \approx$ 40$^{\circ}$ 
for the telescopes, was motivated by the
aim of a good coverage at mid-rapidity.

Additional detectors were employed in order to probe the charged-particle
multiplicity for impact parameter selection. The angular range of
$\theta_{lab}$ = 6$^{\circ}$ to 20$^{\circ}$ was covered by an
azimuthally symmetric array of 36 CaF$_2$-plastic phoswich detectors 
\cite{rabe}. Within the angular range 30$^{\circ}$ to 55$^{\circ}$,
a solid angle of 0.7 sr was covered by an array of 48 elements
of Si-strip detectors. The results presented in the following were 
obtained after selecting an event class of highest associated multiplicity,
corresponding to about 10\% of the total reaction cross section.

\begin{figure}
\centerline{\epsfig{file=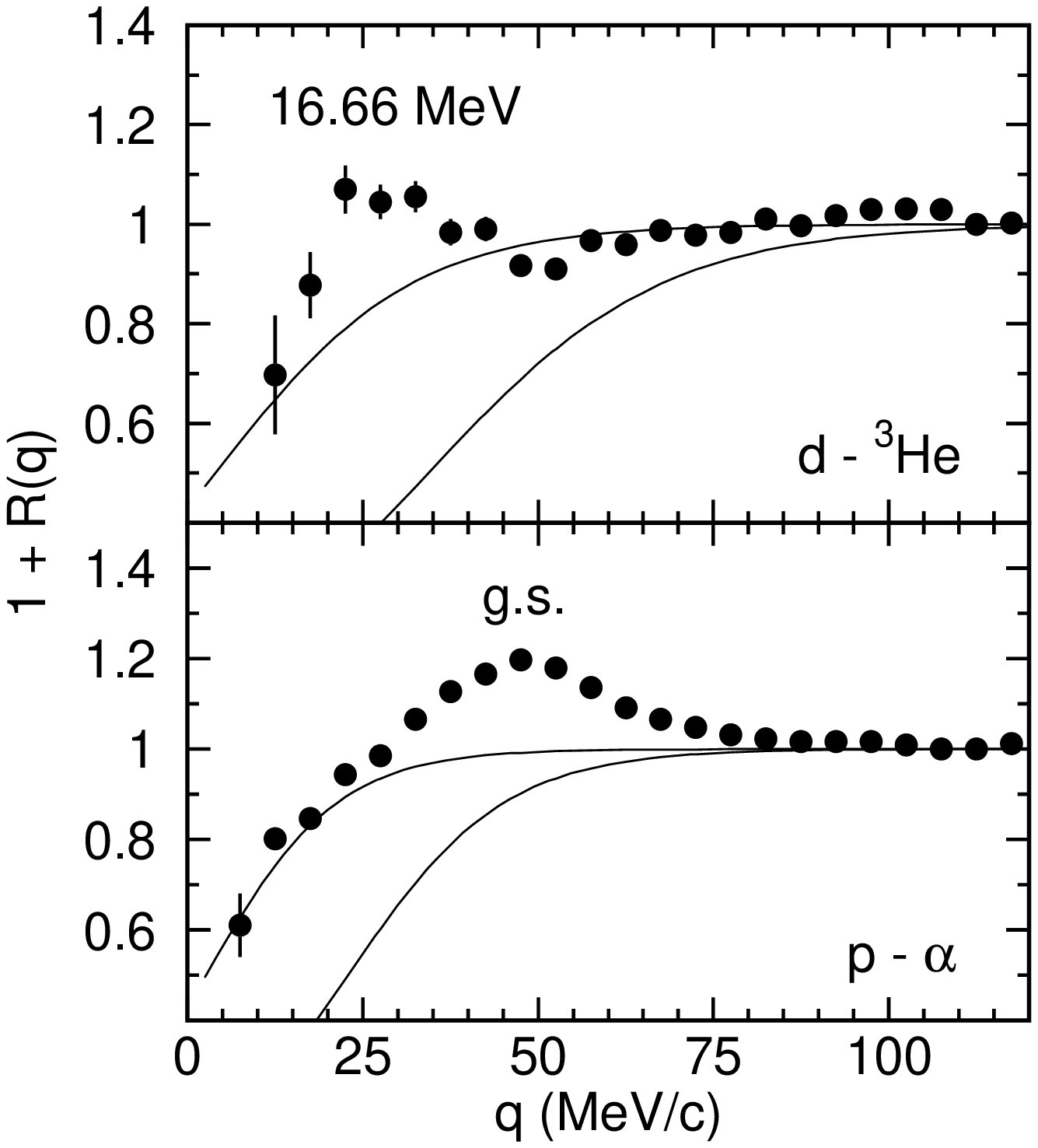,width=\linewidth}}
\caption{
Correlation functions constructed for d-$^3$He (top)
and p-$^4$He (bottom) coincidences as a function of the relative momentum
$q$ of the two particles for central collisions at $E/A$ = 150 MeV.
The observed states in $^5$Li are indicated.
The lines represent the adopted upper and lower bounds
of the non-resonant background. 
}
\label{FIG1}
\end{figure}

The populations of particle-unstable resonances were derived from 
two-particle coincidences measured with the Si-CsI hodoscopes.
The peak structures were identified by using the technique of correlation 
functions, and background corrections were based on results obtained for 
resonance-free pairs of fragments with $Z \le$ 3, 
such as p-d, d-d, up to $^3$He-$^7$Li. Examples of
correlation functions constructed for p-$^4$He and d-$^3$He coincidences
are shown in Fig.~\ref{FIG1}. They are dominated by the resonances corresponding
to the ground state (g.s.) and 16.66-MeV excited state of $^5$Li.
This pair of states represents a widely used thermometer for nuclear 
reactions \cite{poch87,kunde91,schwarz93}.
The observed weak peak intensities are expected for large source sizes. 
Correlated yields of p-t, d-$^4$He, $^4$He-$^4$He, and p-$^7$Li 
coincidences and $^4$He singles yields were also measured and
used to deduce temperatures from the
populations of states in $^4$He (g.s.; group of three states at 
20.21 MeV and higher), $^6$Li (2.19 MeV; group of two states at 4.31 
and 5.65 MeV), and $^8$Be (g.s.; 3.04 MeV; group of five states 
at 17.64 MeV and higher).
The probabilities for the coincident detection 
of the decay products of these resonances were calculated with a
Monte-Carlo model \cite{kunde91,serf97}. The 
uncertainty of the background subtraction is the main 
contribution to the errors of the deduced temperatures.

\begin{figure}
\centerline{\epsfig{file=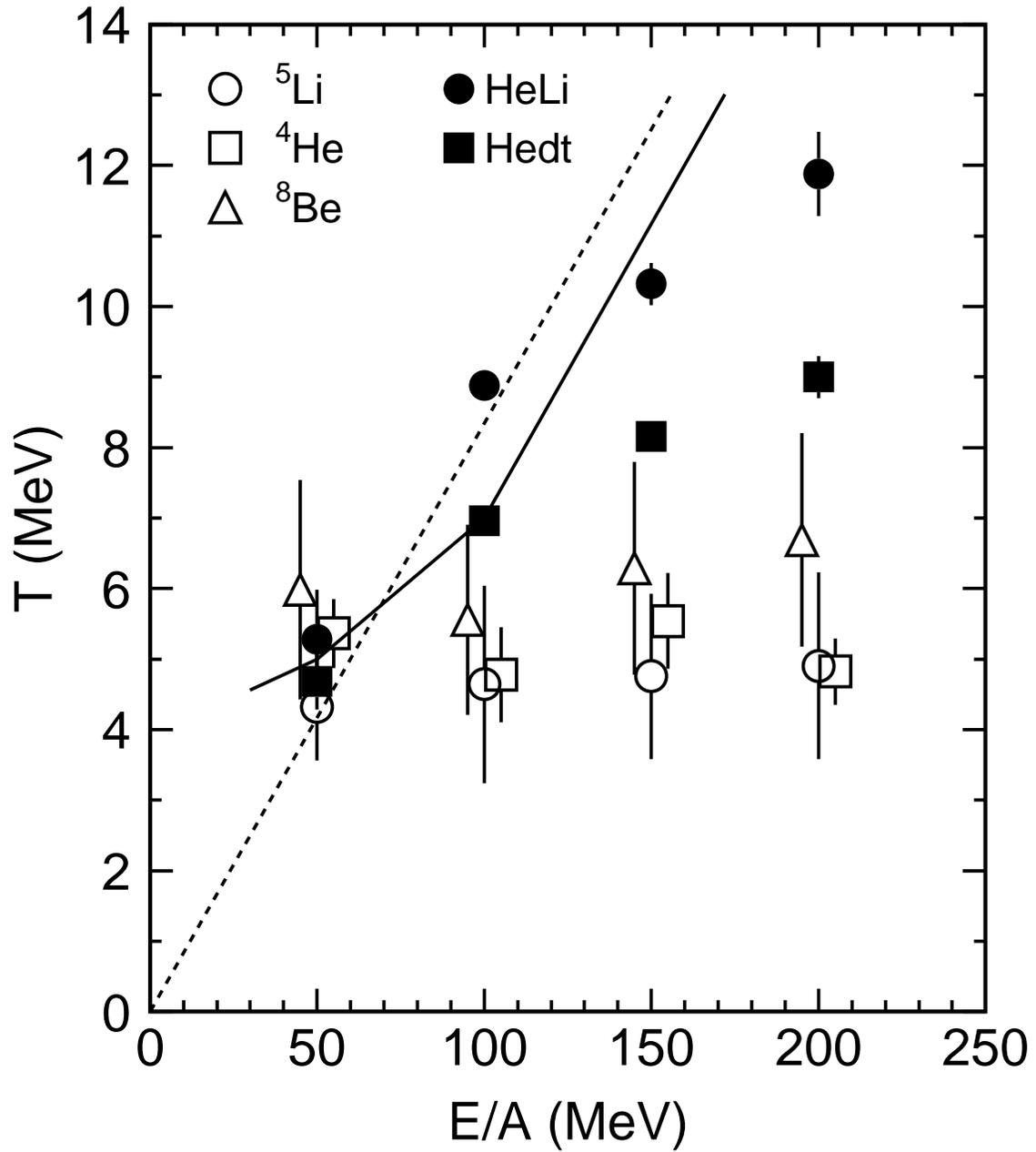,width=\linewidth}}
\caption{
Measured isotope temperatures (full symbols) and excited-state temperatures
(open symbols) as a function of the incident energy per nucleon. The 
indicated uncertainties are mainly of systematic origin. The meaning of the
lines is explained in the text.
}
\label{FIG2}
\end{figure}

The obtained values for two isotope and three
excited-state temperatures are given in Fig.~\ref{FIG2}. 
The isotope temperatures $T_{\rm HeLi}$ and $T_{\rm Hedt}$ 
were derived as described in 
\cite{hongfei}, and the correction factors given there have been applied 
in order to account for the effects of sequential feeding.
The three excited-state temperatures are characterized by large energy 
differences
of the considered states (for $T_{{\rm Be8}}$ the 
18 MeV/ 3.04 MeV result is shown), 
and no corrections for sequential feeding 
were applied (fully justified only for $^5$Li, see below and 
\cite{poch87,schwarz93}).
At $E/A$ = 50 MeV, all temperature values coincide within the interval
$T$ = 4 to 6 MeV, an observation made also at $E/A$ = 35 MeV 
by Huang {\it et al.} \cite{huang97}.
With increasing bombarding energy, however, the isotope
temperatures rise approximately linearly up to
$T_{{\rm HeLi}} \approx$ 12 MeV and $T_{{\rm Hedt}} \approx$ 9 MeV
at $E/A$ = 200 MeV.
The excited-state temperatures, on the other hand, mutually consistent with 
each other, appear to be virtually independent of the bombarding energy. 
Their mean values, over the covered range of bombarding energies, are
4.6 $\pm$ 0.6 MeV, 5.1 $\pm$ 0.3 MeV, and 6.1 $\pm$ 0.7 MeV 
for $T_{{\rm Li5}}$, $T_{{\rm He4}}$,
and $T_{{\rm Be8}}$, respectively. These differences may be significant,
and could even be enhanced by sequential-decay corrections (see below),
but they seem marginal in comparison to the apparent qualitative 
difference between the isotope and excited-state temperatures.

\begin{figure}
\centerline{\epsfig{file=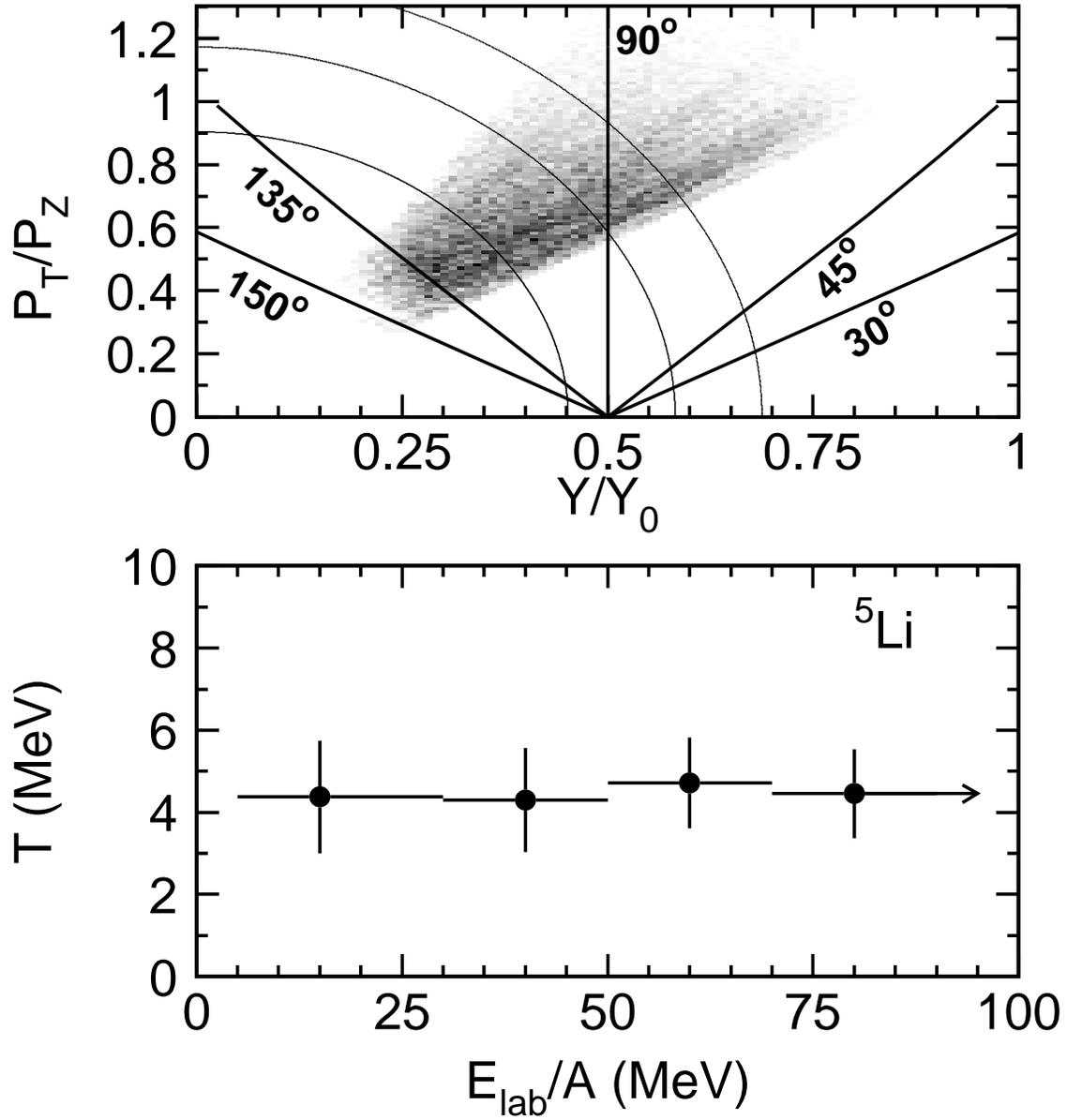,width=\linewidth}}
\caption{
Top: Acceptance of the 96-element hodoscope for p-$^4$He coincidences with
relative momenta 25 $\le q \le$ 75 MeV/c in the plane subtended by relative 
transverse momentum and normalized rapidity of the corresponding 
$^5$Li fragment for $E/A$ = 150 MeV.
The loci of $\Theta_{cm}$ = 30$^{\circ}$, 45$^{\circ}$, 90$^{\circ}$, 
135$^{\circ}$, and 150$^{\circ}$ (thick lines) and of $^5$Li energies 
$E_{lab}/A$ = 30, 50, and 70 MeV (thin lines) are indicated.\\
Bottom: Temperatures derived from the $^5$Li ground and excited states for the
four intervals of the $^5$Li laboratory energy. Note that the errors are
mainly of systematic origin.
}
\label{FIG3}
\end{figure}

The momentum-space acceptance of the detectors, kept at fixed positions
in the laboratory, changes with bombarding energy in the center-of-mass 
frame. For the case of $E/A$ = 150 MeV, the 
acceptance of the 96-element hodoscope for p-$\alpha$ coincidences in the 
momentum interval corresponding to the $^5$Li-g.s. resonance
is shown in Fig.~\ref{FIG3} (top). It covers the region around 
$\theta_{cm} = 90^{\circ}$ and, in addition, extends to 
forward and backward angles with a varying transverse-momentum acceptance.
The wide acceptance and its shift with bombarding energy 
should not be crucial, however, because no significant variation of 
$T_{{\rm Li5}}$ within the
covered momentum space was found (Fig.~\ref{FIG3}, bottom).

It is not immediately obvious that the divergence of the
isotope and excited-state temperatures, growing dramatically with 
bombarding energy, contradicts the concept of a common fragment
freeze-out at a single temperature. Xi {\it et al.} report that their 
recent statistical calculations indicate a strongly reduced 
sensitivity of the helium-lithium thermometer at high temperature, 
such that it may prevent reliable temperature measurements at 
$T >$ 7 MeV \cite{xi96}. 
Accordingly, a consistent common temperature, if existing, should be low.
The excluded-volume effect, as incorporated in the
quantum-statistical model by Gulminelli and Durand
\cite{gulm97}, causes a suppression of particle-unstable resonances 
decaying into loosely bound products, such as the 16.66-MeV excited state 
of $^5$Li. It will have the effect that the apparent 
$T_{{\rm Li5}}$ is low while
a common emission temperature may be high (cf. \cite{ass97}).
These calculations demonstrate that
large effects can be caused by sequential decay and by structural
differences of the nuclear states employed in the temperature measurements,
even though they may not suffice to give a consistent explanation of all
the present observations.

The dynamical evolution of the fragment formation has very recently been 
investigated with
transport models \cite{dani92}, including 
nuclear molecular-dynamics \cite{barz96,strach97} and
quantum molecular-dynamics \cite{dorso95,puri96} models, 
applied to the present and similar reactions.
These studies suggest that the asymptotic fragments can be 
identified at an early stage of the reaction, typically at $\approx$ 40 fm/c.
These times coincide with the development of the collective flow component
of the fragment motion \cite{dani92,barz96,strach97}. If local chemical 
equilibrium has been reached the isotopic composition should 
reflect the temperature of the system at that particular time.

According to various flow analyses, between 40\% and 60\% of the 
collision energy is converted into collective flow energy \cite{reis97}.  
In the simplest approximation, the breakup temperature is then
estimated as $T = (E/A)/12$. This assumes complete stopping
of the incident nuclei and a classical gas with $3 \cdot 2A$ degrees of
freedom carrying a thermal energy component of 50\% of the
collision energy.
This relation (dashed line in Fig.~\ref{FIG2}) does not describe the
data very well, but it illustrates the expected linear rise and 
shows that the measured isotope temperatures 
have about the right order of magnitude. 
Better agreement with the data at the lower energies is obtained 
if, for the same thermal energies, the experimental temperature vs. energy
relation of Ref. \cite{poch95} is used (Fig.~\ref{FIG2}, full line).
Even though it remains to be understood why $T_{{\rm Hedt}}$ is 
considerably lower in the present case (cf. \cite{hongfei,haug96,ma97}),
the comparison suggests
that the isotope thermometers are sensitive to the local temperature 
at freeze-out in a blast scenario \cite{bond78,siemens,stoe81}.

The excited states used for the temperature evaluation are very specific
quantum states with widths of 1 MeV or less. They are unlikely
to exist in the nuclear medium in identical forms \cite{dani92,roepke,alm95}.
The observed asymptotic states can develop or survive only
at very low densities that may not be reached before 
the cluster is emitted into vacuum. 
Accordingly, the excited-state populations should reflect the 
temperature and its fluctuations at this final stage of fragment emission.
The molecular dynamics calculations show that a cluster continues to
interact with the surrounding cooling and expanding matter for a 
considerable time after it has been formed \cite{barz96}. 
This will lower
its internal excitation but, apparently, does not change as much the
isotopic composition.

\begin{figure}
\centerline{\epsfig{file=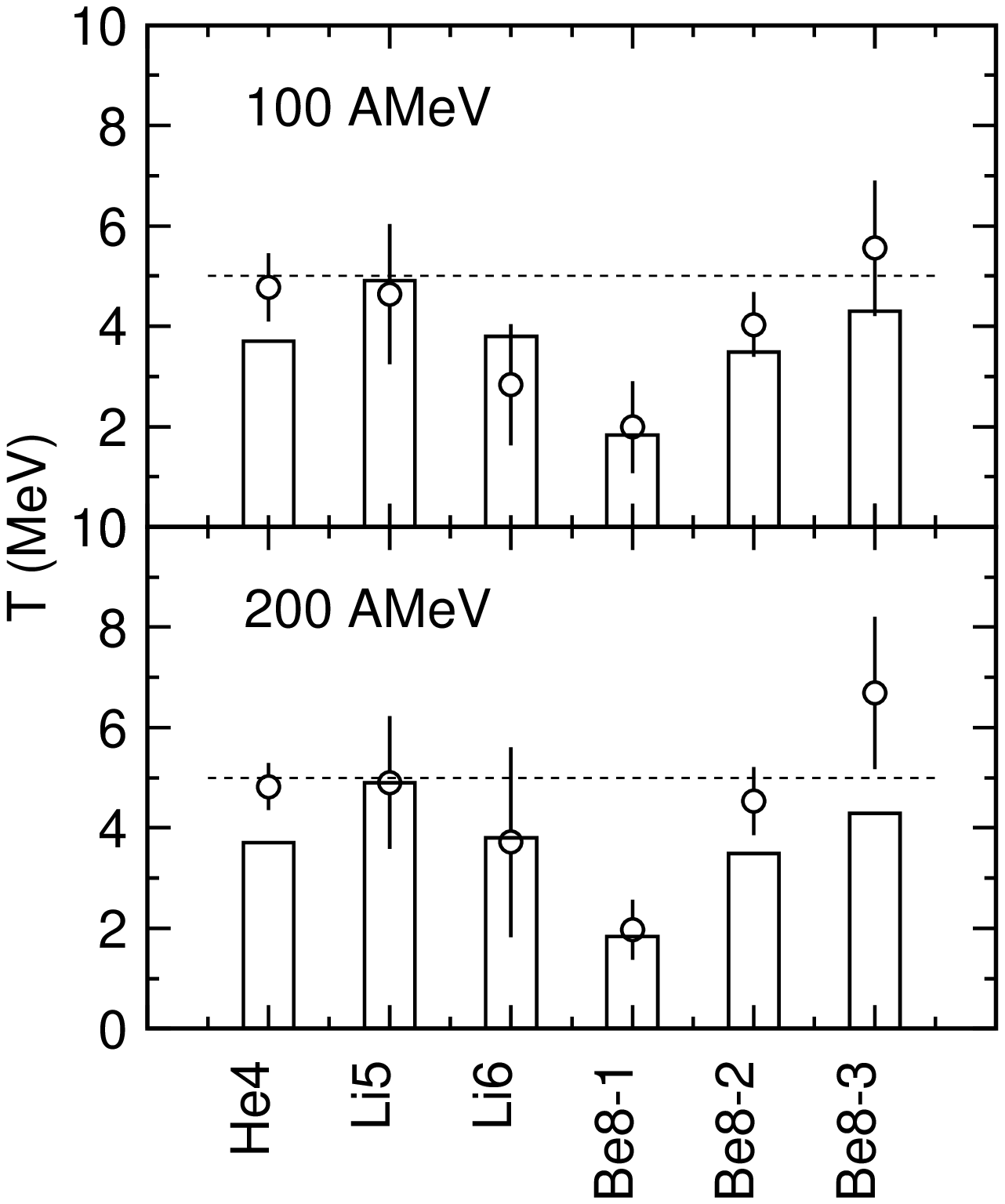,width=\linewidth}}
\caption{
Six values of apparent excited-state temperatures deduced from states 
in $^4$He,
$^5$Li, $^6$Li, and $^8$Be (open circles) in comparison to predictions of 
the quantum-statistical model (bars) for E/A = 100 MeV (top) and 200 MeV
(bottom). The calculations were performed with $T$ = 5 MeV (dashed lines) 
and $\rho / \rho_0$ = 0.1. The notation for $T_{{\rm Be8}}$ is 1: 
3.04 MeV/g.s., 2: 18 MeV/g.s., and 3: 18 MeV/3.04 MeV.  
}
\label{FIG4}
\end{figure}

Excited state populations have thermal characteristics \cite{nayak}
and have been shown to correspond to expected temperatures in compound 
reactions \cite{schwarz93,dabr90}. 
In the present case, the observed internal fragment excitations,
associated with the final breakup of the system,
are found to be consistent with 
a thermal population at $T$ = 5 to 6 MeV (Fig.~\ref{FIG4}).
The apparent temperatures $T_{{\rm Li6}}$ and
$T_{{\rm Be8-1}}$ (3.04 MeV/g.s.),
derived from states not widely separated in energy, are lower but
in accordance with the side-feeding effects predicted by 
the quantum-statistical model \cite{konop94}. 

With the internal excitations corresponding to lower temperatures,
the side-feeding corrections for the isotope temperatures will be 
rather complex. While the role
of highly excited continuum states may be reduced, the corrections 
to be expected may still be large. To give an example, 
for $T_{{\rm HeLi}}$, the
quantum-statistical model predicts a modification of the isotopic
double ratio by a factor of 1.5 at $T$ = 5 MeV. 
It corresponds to a 20\% modification of $T$ at this temperature,
but to larger relative corrections at higher temperatures 
(e.g., 30\% near $T$ = 10 MeV).
Such corrections will have the effect of further increasing the slope of 
the isotope temperatures vs. bombarding energy 
and of improving the agreement 
with the simple expectations (Fig.~\ref{FIG2}).

The presented interpretation of the observed qualitative difference 
between the isotope and excited-state temperatures seems rather attractive. 
It implies that isotope yields may be used to probe the early stages of 
the fragment formation process, and it may explain
the saturation of the excited-state temperatures that characterizes
a wide variety of measurements at intermediate
and relativistic energies \cite{kunde91,schwarz93}. This interpretation,
therefore, should be confirmed by further work which may aim at a
quantitative interpretation of the internal fragment excitation but also 
address current open questions such as the role of initial correlations
(see, e.g., \cite{kunde91,barz96,puri96,knoll88}) 
and of quantum effects in the fragment formation process
\cite{dona94,ono96,ohni97,feld97}). 
 
Extensive discussions with J. Aichelin, H.W.~Barz, 
F. Gulminelli, G. R\"opke, and M.B. Tsang are
gratefully acknowledged.
J.P. and M.B. acknowledge the financial support
of the Deutsche Forschungsgemeinschaft under the Contract No. Po 256/2-1
and Be1634/1-1, respectively.
This work was supported by the European Community under
contract ERBFMGECT950083.

\begin{references}

\bibitem[a]{AAA}
Present address:
Department of Physics, Yale University, New Haven CT 06512, USA

\bibitem[b]{BBB}
Present address:
National Superconducting Cyclotron Laboratory, Michigan State University, 
East Lansing, MI 48824, USA

\bibitem{poch95}
J.~Pochodzalla {\it et al.},
Phys.\ Rev.\ Lett. {\bf 75}, 1040 (1995).

\bibitem{gross90}
D.H.E.~Gross,
Rep.\ Prog.\ Phys. {\bf 53}, 605 (1990).

\bibitem{bond95}
J.P.~Bondorf {\it et al.},
Phys.\ Rep. {\bf 257}, 133 (1995).

\bibitem{hongfei}
Hongfei Xi {\it et al.}, 
Z. Phys. {\bf A}, in press.

\bibitem{xi96}
Hongfei Xi {\it et al.},
Phys.\ Rev. {\bf C54}, R2163 (1996).

\bibitem{gulm97}
F.~Gulminelli and D.~Durand,
Nucl.\ Phys. {\bf A615}, 117 (1997).

\bibitem{natowitz}
J.B.~Natowitz {\it et al.},
Phys.\ Rev. {\bf C52}, R2322 (1995).

\bibitem{more96}
L.G.~Moretto {\it et al.},
Phys.\ Rep. {\bf 287}, 249 (1997).

\bibitem{mori94}
D.J.~Morrissey {\it et al.},
Ann.\ Rev.\ Nucl.\ Part.\ Science {\bf 44}, 65 (1994).

\bibitem{reis97}
W.~Reisdorf {\it et al.},
Nucl.\ Phys. {\bf A612}, 493 (1997),
and references therein.

\bibitem{albergo}
S.~Albergo {\it et al.},
Il Nuovo\ Cimento {\bf 89A}, 1 (1985).

\bibitem{poch87}
J.~Pochodzalla {\it et al.},
Phys.\ Rev. {\bf C35}, 1695 (1987).

\bibitem{kunde91}
G.J.~Kunde {\it et al.},
Phys.\ Lett. {\bf B272}, 202 (1991).

\bibitem{rabe}
H.J. Rabe {\it et al.},
Phys.\ Lett. {\bf B196}, 439 (1987).

\bibitem{schwarz93}
C.~Schwarz {\it et al.},
Phys.\ Rev. {\bf C48}, 676 (1993).

\bibitem{serf97}
V.~Serfling, 
PhD thesis, Universit\"at Frankfurt, 1997, unpublished.

\bibitem{huang97}
M.J.~Huang {\it et al.},
Phys.\ Rev.\ Lett. {\bf 78}, 1648 (1997).

\bibitem{ass97}
M.~Assenard {\it et al.},
preprint SUBATECH 97-15 (1997).

\bibitem{dani92}
P.~Danielewicz and Q.~Pan,
Phys.\ Rev. {\bf C46}, 2002 (1992).

\bibitem{barz96}
H.W.~Barz {\it et al.},
Phys.\ Lett. {\bf B382}, 343 (1996).

\bibitem{strach97}
A.~Strachan and C.O.~Dorso,
Phys.\ Rev. {\bf C55}, 775 (1997).

\bibitem{dorso95}
C.O.~Dorso and J.~Aichelin,
Phys.\ Lett. {\bf B345}, 197 (1995).

\bibitem{puri96}
R.K.~Puri {\it et al.},
Phys.\ Rev. {\bf C54}, R28 (1996).

\bibitem{haug96}
J.A.~Hauger {\it et al.},
Phys.\ Rev.\ Lett. {\bf 77}, 235 (1996).

\bibitem{ma97}
Y.-G.~Ma {\it et al.},
Phys.\ Lett. {\bf B390}, 41 (1997).

\bibitem{bond78}
J.P.~Bondorf {\it et al.},
Nucl.\ Phys. {\bf A296}, 320 (1978).

\bibitem{siemens}
P.J.~Siemens and J.O.~Rasmussen,
Phys.\ Rev.\ Lett. {\bf 42}, 880 (1979).

\bibitem{stoe81}
H.~St\"ocker {\it et al.},
Z. Phys. {\bf A 303}, 259 (1981).

\bibitem{roepke}
M.~Schmidt {\it et al.},
Ann.\ Phys. (N.Y.) {\bf 202}, 57 (1990).

\bibitem{alm95}
T.~Alm {\it et al.},
Phys.\ Lett. {\bf B346}, 233 (1995).

\bibitem{nayak}
T.K.~Nayak {\it et al.},
Phys.\ Rev.\ Lett. {\bf 62}, 1021 (1989).

\bibitem{dabr90}
H.~Dabrowski {\it et al.},
Phys.\ Lett. {\bf B247}, 223 (1990).

\bibitem{konop94}
J. Konopka {\it et al.},
Phys.\ Rev. {\bf C50}, 2085 (1994).

\bibitem{knoll88}
J. Knoll and J. Wu,
Nucl.\ Phys. {\bf A481}, 173 (1988).

\bibitem{dona94}
R. Donangelo {\it et al.},
Phys.\ Rev. {\bf C50}, R563 (1994).

\bibitem{ono96}
A. Ono and H. Horiuchi,
Phys.\ Rev. {\bf C53}, 2958 (1996).

\bibitem{ohni97}
A. Ohnishi and J. Randrup,
Phys.\ Lett. {\bf B394}, 260 (1997).

\bibitem{feld97}
H.~Feldmeier and J.~Schnack,
Prog.\ Part.\ Nucl.\ Phys. {\bf 39}, 393 (1997).

\end{references}
\end{document}